\documentclass[11pt]{iopart}

\usepackage{iopams}  
\usepackage{bm, mathrsfs, dsfont, braket, graphicx,array}
\usepackage[latin1]{inputenc}
\usepackage[enableskew]{youngtab}

\def\ie{{\it i.e.},\ }
\def\ig{{\it i.g.},\ }
\def\eg{{\it e.g.}\ }

\newcommand{\nn}{\nonumber}
\newcommand{\vc}[1]{\bm{{#1}}}
\newcommand{\vco}[1]{\hat{\bm{#1}}}
\newcommand{\myO}[1]{\mathrm{O}\left(#1\right)}
\newcommand{\z}{\mathrm{z}}
\newcommand{\tot}{\mathrm{tot}}

\graphicspath{{./}{./pics/}{./submit/pics/}}

\begin{document}

\title{Momentum classification of SU($\bm{n}$) spin chains using extended Young Tableaux}

\author{Burkhard Scharfenberger$^1$ and Martin Greiter$^2$}
\address{$^1$Institut f\"ur Theorie der Kondensierten
  Materie, KIT, Campus S\"ud,\\ D-76128 Karlsruhe} 
\address{$^2$Lehrstuhl f\"ur Theoretische Physik 1, Fakult\"at f\"ur Physik und Astronomie, 
Julius-Maximilians-Universit\"at, Am Hubland, D-97074 W\"urzburg} 

 \pagestyle{plain} 
 \date{\today}


\begin{abstract}
  Obtaining eigenvalues of permutations acting on the product space of
  $N$ representations of SU($n$) usually involves either diagonalising
  their representation matrices on total-weight subspaces or
  decomposing their characters, which can be obtained from Frobenius'
  formula or via graphical methods using Young tableaux. 
  For products of fundamental representations of SU($n$), Schuricht and
  one of us proposed the method of extended Young Tableaux,
  which allows reading the eigenvalues of the cyclic permutation $C_N$
  directly off the, slightly modified, standard Young tableaux labelling an
  irreducible SU($n$) representation. 
  Here we generalise the method to all symmetric representations of SU($n$), 
  and show that  $C_N$ eigenvalue computation based on extended
  Young tableaux is  at least linearly faster than the standard methods
  mentioned.
\end{abstract}

\maketitle

\section{Introduction \label{sec:intro}}
Symmetries, whether discrete or continuous, have been a central concept in physics 
since its earliest days, and only by making use of them, explicitly or implicitly \eg by 
choosing a suitable coordinate system, can most systems be treated analytically or 
even numerically. 
This is also true in the study of spin lattice models, a vibrant field of contemporary
condensed matter physics that has produced many insights into novel states of matter 
and manifestations of order. 
A spin model consists of a cluster of $N$ spins arranged on some lattice tile, usually 
with periodic, but possibly other, boundary conditions, and a Hamiltonian describing 
the interaction of the spins with each other or with external fields. 
The 'spins' transform like some (irreducible) representation of SU($n$), usually 
SU($2$), and thus the Hamiltonian acts on a tensor product space whose dimension grows 
exponentially in $N$. In recent years, however, models with higher $n$ have also received 
attention~\cite{sutherland75prb3795,affleck88npb582,affleck-91npb467,kawakami92prb1005,
kawakami92prb3191,ha-93prb12459,shen01prb132411,honerkamp-04prl170403,
damerau-07jsm1204,greiter-07prb184441,pankov-07prb104436,arovas08prb104404,
katsura-08jpa135304,xu-08prb134449,manmana-11pra043601,hermele-11prb174441,
 corboz-11prl215301,bauer-12prb125116},
especially since  cold atoms in optical lattices hold the prospect of realising SU($n$) models
experimentally~\cite{ripoll-04prl250405,hermele-09prl135301,azaria-09pra041604,xu10prb144431,
gorshkov-10np289}.

The most commonly considered interaction is of the Heisenberg form 
$H=\sum_{i<j} J_{ij}\vco{S}_i\vco{S}_j$ 
between spins on sites $i$ and $j$ with a coupling  constant $J_{ij}$. 
Such a Hamiltonian is inherently invariant under global SU($2$) (SU($n$)) rotations and 
conserves both $S_{\tot}$ and $S^\z_{\tot}$ (or, for general SU($n$), highest total weight 
$\bm{w}_\tot$ and total weight $\bm{w}_\tot^\z$) respectively.
Usually the $J_{ij}$ obey some symmetry relations, often they even possess the full 
symmetry of the underlying lattice, implying that the Hamiltonian is  conserved under 
all operations in $\mathscr{L}$, the (point) symmetry group of  the lattice. 
Since in numerical studies the lattice is some finite tile  containing $N$ sites, 
$\mathscr{L}$ is a subgroup of $S_N$, the group of all permutation of N objects.

In treating such a spin lattice model, either analytically or numerically, it seems clear 
that one should exploit the symmetries of the problem as far as possible. 
Therefore a product basis, where the z-components of all individual spins provides 
a complete labelling of all states, while conceptually simple, is not the best choice from 
a performance perspective. Rather, we should use a basis labelled by
the eigenvalues of a maximal commuting subset of $\mathscr{L}$ plus 
a number of other labels, \eg the eigenvalues of as many further commuting 
permutations from $S_N$ as are needed to provide a unique labelling.

A mathematical problem that arises in this context is to determine the eigenvalues 
of these labelling lattice symmetries. 
Changing language from SU($2$) to SU($n$) the general problem can be stated 
like this: for arbitrary $N$-fold product spaces $V_{\bm{v}}^{\otimes N}$ of some 
irreducible representation (irrep) $V_{\bm{v}}$ of SU($n$), find the eigenvalues 
the labelling symmetries $L$.
The ($n$-1)-dimensional vector $\bm{v}$ is the highest weight of the irrep and has 
the same meaning for $SU(n)$ as spin for SU(2).

There are two traditional ways to solve this computationally.
we know that the tensor product  decomposes into a 
direct sum of irreducible representations of SU($n$):
\begin{equation}
 V_{\bm{v}}^{\otimes N} = \bigoplus_{\bm{w}} V_{\bm{w}}^{\oplus a_{\bm{w}}}
 \label{eq:proddecomp}
\end{equation}
Say we want to know the eigenvalues for our labelling lattice symmetries on 
the subspace $V_{\bm{w}}^{\oplus a_{\bm{w}}}$ 
of all irreps $V_{\bm{w}}$ of SU($n$) (\ie $V_{\bm{w}}$ appears $a_{\bm{w}}$ 
times in  $V_{\bm{v}}^{\otimes N}$). 
Then one way is to use character theory to determine the irreducible representations 
of $S_N$ contained in $V_{\bm{w}}^{\oplus a_{\bm{w}}}$.  
The eigenvalues of the permutations are then obtained 
by diagonalising their representation matrices in these irreps, which is 
possible for all of them simultaneously since they commute.
Alternatively, we could simply write down all product states $\phi_{\bm{w}}
\in V_{\bm{v}}^{\otimes N}$, which have the total weight $\bm{w}^\z_\tot=\bm{w}$ 
(in the case of SU(2) this corresponds to a fixed $S^\z_\tot$ subspace),
determine the representation matrices of the labelling symmetries 
and again diagonalise them (all simultaneously).
In both cases, we need to repeat the process for those highest
weight multiplets $\bm{w'}>\bm{w}$, that are contained in the subspace of the
total weight $\bm{w}^\z_\tot=\bm{w}$. In the case of SU(2) for instance, 
it is sufficient to consider the next higher $S^\z_\tot$ subspace, 
\ie $S^\z_\tot=S\!+\!1$, and disregard all sets of eigenvalues which 
appear in both subspaces.

Young tableaux are a diagrammatic technique originally invented to compute various 
properties of irreps of the permutation group $S_N$~\cite{young1}, 
but they and other techniques based on them have since seen a myriad of uses in
both mathematics~\cite{hamermesh89, fulton-91} and physics~\cite{wu-92,yu-96tca125}.
Their usefulness is mostly related to the Schur-Weyl duality, which makes a direct connection 
between irreps of GL($n$) and $S_N$ contained in tensor products of some elementary 
GL($n$) irrep. Due to this, YT also provide an elegant means of obtaining the decomposition of 
$V_{\bm{v}}^{\otimes N}$ into irreducible representations of SU($n$),
 \ie of obtaining the $a_{\bm w}$ in~(\ref{eq:proddecomp}).

A further use of YT was introduced in a 2007 paper~\cite{greiter-02jltp1029} 
(see also~\cite{greiter11}), 
by Schuricht and one of us: the method of extended Young tableaux.
It allows the spinon content of an eigenstate of the Haldane-Shastry-Model for a 
chain of $N$ fundamental SU($n$) spins to be read off directly from slightly 
modified Young tableaux. 
The interesting consequence we want to point out here, is that this also allows 
one to find the state labels $\ket{\bm{w}_\tot, p}$ for the symmetry group of the 
1D chain. 
This group is the cyclic group $\mathscr{C}_N$ generated by the single-site 
translation $C_N$ where each eigenvalue $\gamma$ of $C_N$  can be identified 
with a momentum $p$ along the chain via the equation
$\gamma=\exp[\mathrm{i}p]$. 
To our knowledge, this is the only such method working directly with 
Young tableaux and as a practical consequence, this enables a 
significant speed-up of $C_N$-eigenvalue computations.

In this paper we generalise the method of extended Young tableaux to 
higher representations of SU($n$) and present numerical evidence that it
indeed gives the correct results for the eigenvalues of $C_N$, as long
as the representations $V_\sigma$ which are coupled are
\emph{symmetric}, \ie correspond to Young tableaux with a single row. 
We also find, that while it does give the correct distribution of momenta 
on the subspace $V_\lambda^{\oplus a_\lambda}$ of all multiplets $\lambda$, 
it does not assign these momenta to the individual YT in a way that would 
allow deducing the irreducible $S_N$-representation content of 
$V_\lambda^{\oplus a_\lambda}$. A positive result would both have been
quite useful in itself and would also have given a physical meaning
to individual YT, similar in spirit to the connection between YT and angular
momentum states sought by McAven and Schlesinger~\cite{mcaven-01jpa8333}.
Lastly we show how working directly with the Young tableaux of an irreducible 
representation speeds up the computation of $C_N$ eigenvalues over traditional 
methods by at least a linear factor $N$.

This paper is organized as follows: In section 2 we briefly review
the method of Young tableaux for $S_N$ and how it relates to the
irreducible representations of SU($n$). In section 3, we restate
the extension procedure for fundamental representations and reformulate
it in a way better suited to both numerical implementation and generalisation 
to product spaces of higher representations. We complete this task 
in section 4, which contains the main result, namely that the procedure appears 
to work for \emph{all symmetric higher SU($n$) representations}. 
Furthermore, we comment on relation between the extended Young tableaux 
method and irreducible representations of $S_N$ contained in the product 
spaces. 
In section 5 we compare the computational complexity of the extension 
procedure to that of generic methods of obtaining the eigenvalues of 
$C_N$. 
Finally, we summarize our results in section 6.

\section{Young Tableaux and SU($n$)\label{sec:ytsun}}
We begin with a short summary of Young tableaux and some of their traditional uses.

\begin{figure}
 \centering
 \includegraphics[width=.51\textwidth]{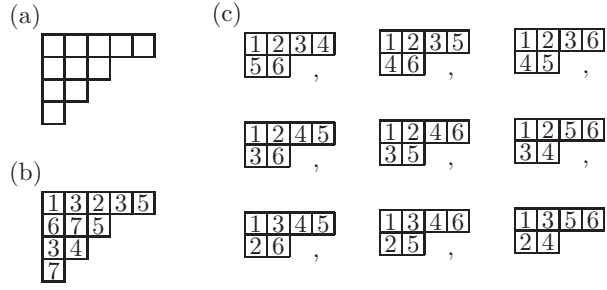}
 \caption{(a) the Young diagram or shape to the partition $(5,3,2,1)$, 
          (b) the same diagram as a (general) Young tableaux
          (c) all standard Young tableaux on the shape $(4,2)$}
 \label{fig:YD}
\end{figure}

A \emph{Young diagram} or \emph{shape} is a graphical depiction of an integer 
partition
\begin{eqnarray}
 && (\lambda)=(\lambda_1,\lambda_2,\dots,\lambda_k),\;\lambda_1\geq\lambda_2\geq\dots\geq\lambda_k>0, \nn\\
 && |\lambda| := \sum_j \lambda_j = N,
\end{eqnarray}
as $k$ left-justified rows of boxes, where row $j$ has $\lambda_j$ boxes in it.
A \emph{Young tableau} (YT) on the shape $\lambda$ is any filling of the boxes with
integers between $1\dots N$ (see Fig.~\ref{fig:YD}a and 1b ).
Counting Young tableaux subject to certain building rule is what is at the
heart of their application in the representation theory of both the symmetric
group $S_N$, the group of all permutations of $N$ distinguishable things, and
SU($n$), the group of special unimodular, complex $n\times n$ matrices.

\emph{$S_N$. ---}
The irreducible representations of the symmetric group $S_N$ can be labelled by
integer partitions $\lambda,\,|\lambda|=N$. Originally, Young tableaux where 
invented to provide a 
graphical method of computing the character $\chi^{\lambda}(P)$ of an element 
$P\in S_N$ in the irreducible representation $\lambda$~\cite{young1}. 
Of special significance is the character of the identity, since it is equal the dimension of an 
irreducible representation: $\chi^{\lambda}(\textrm{id})=\textrm{dim}(\lambda)=:N_\lambda$. 
This dimension can be determined by counting all \emph{standard Young tableaux} 
on the shape $\lambda$. A tableau is called standard if the numbers in its boxes are 
strictly increasing in both rows and columns.
Fig.\ref{fig:YD}c) for instance shows all standard YT on $\lambda=(4,2)$. There
are 9 of them, therefore the representation $(4,2)$ of $S_6$ must be 9 dimensional.

\begin{figure}[b]
  \centering
  \includegraphics[width=.534\textwidth]{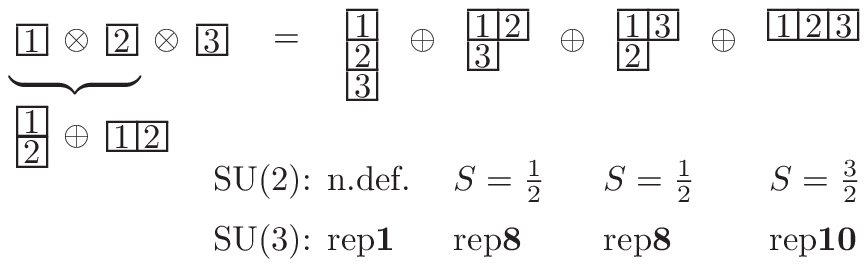}
  \caption{Building higher irreducible representations from the fundamental one
                 via branching in the case of SU(2) and SU(3). Counting dimensions
                 as a check, we see that the decompositions are complete in both cases:
                 $2^3=2+2+4$ and $3^3=1+8+8+10$.}
  \label{fig:ytfig_branch}
\end{figure}

\emph{Special unitary group SU($n$) ---}
In the context of SU($n$) Young tableaux can be applied to decompose tensor products 
of representations. An irrep of SU($n$) is characterized by its highest weight, where a 
weight is the ($n$-1)-dimensional vector of eigenvalues of the simultaneously diagonalisable 
group generators spanning the ($n$-1)-dimensional Cartan sub-algebra of su($n$), the 
generating Lie algebra of SU($n$). 
A weight $\bm{w}=(w_1,w_2,\dots,w_{n-1})$ is said to be higher than a weight 
$\bm{w'}=(w'_1,\dots,w'_{n-1})$ if the first nonzero entry in $\bm{w}-\bm{w'}$ is positive. 
In the well known case of SU(2), for instance, irreducible representations are characterized 
by their spin $S$, which can be integer or half-integer ($S=\frac{1}{2},1,\frac{3}{2},\dots$),
and the highest weight corresponds simply to the highest possible value of $S^\z$, which
is $S^\z = S$. 
Since SU($n$) is defined as a matrix group, one representation is always the group itself,
its carrier vector space being $\mathds{C}^n$. It is irreducible and of special interest, because 
by forming tensor products of multiple $\mathrm{F}_n$ and projecting onto subspaces of 
appropriate symmetry we can form all irreducible representations of SU($n$),  This is the main 
consequence of the the already mentioned Schur-Weyl-duality.

Thus, Schur-Weyl duality in effect implies that one can use Young tableaux to decompose 
tensor products of $\mathrm{F}_n$ (or indeed higher SU($n$) representations).
Associating $\mathrm{F}_n$ with a single box Young diagram, there is a neat diagrammatic
way to do this: we simply construct all $N$-box standard Young tableaux with no more than 
$n$ rows, which can best be done using the branching rule (see \eg~\cite{hamermesh89}). 
The process is illustrated in Figure~\ref{fig:ytfig_branch}. 
This also means there is a 1-1 correspondance between an $n\! -\! 1$ dimensinoal heighest
weight $\bm{w}$ and an $n\! -\! 1$-row shape $\lambda$, so we can from now on use
shapes not only to index $S_N$- but also SU($n$)-irreps.

\begin{figure}[t]
  \centering
     \includegraphics[width=.596\textwidth]{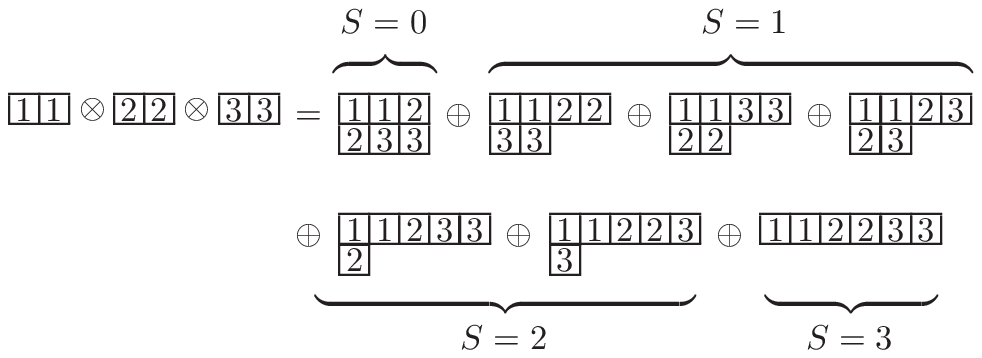}
 \caption{Decomposition of a $3\times(S=1)$ product space of SU(2)}
 \label{fig:ytfig_N3S1}
\end{figure}

It is possible to generalise this procedure to decompose the product space $V_\sigma^{\otimes N}$ 
of arbitrary irreducible representations associated with the shape $\sigma=(\sigma_1,\dots,\sigma_l)$. 
In the case of single-row $\sigma$ ('\emph{symmetric representation}', $l=1$), which is all we will 
need in this paper, this generalisation is straightforward: We again use the branching rule, \ie add 
boxes step by step, but now each number $j=1,\dots,N$ appears $|\sigma|$ times instead of only 
once and we have to take care not to put two boxes with the same number on top of each other 
(see Fig.~\ref{fig:ytfig_N3S1}).

As we have stated, the number $N_\lambda$ of standard YT on the shape $\lambda$ equals both 
the dimension of the irreducible $S_N$ representation labelled with $\lambda$ and the multiplicity 
of the irreducible SU($n$) representation associated with $\lambda$ (via the correspondence 
highest weight $\leftrightarrow$ Young diagram) tensor product $F_n^{\otimes N}$ of the 
fundamental representation of SU($n$).
This just another consequence of the Schur-Weyl-duality: one can show that the subspace 
$V_\lambda^{N_\lambda}$ in the tensor product $\mathrm{F}_n^{\otimes n}$  always forms 
an irreducible representation of the permutation group $S_N$ equivalent to the irrep labelled 
by the integer partition $\lambda$.

For an example we can again consider Figure~\ref{fig:ytfig_branch}, the two YT of shape $(2,1)$ 
tell us, as mentioned, that $F_2^{\otimes 3}$ contains two doublets but also, that the states of 
these doublets (for each fixed $S^\z_\tot$) transform like a standard representation (2,1) under 
the action of the group $S_3$.

\section{Extended Young tableaux\label{sec:eYT}}

We now turn to the problem described in the introduction of computing the eigenvalues of 
the cyclic permutation $C_N$ on total spin subspaces $V_\lambda^{\oplus N_\lambda}$.
Let us first review the extension procedure for YT of fundamental representations 
introduced in~\cite{greiter-02jltp1029}.

\begin{figure}[b]
  \centering
  \includegraphics[width=.46\textwidth]{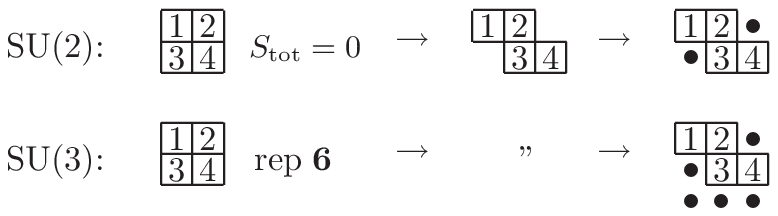}
  \caption{A simple example of the extension procedure: the lower row
    slides to the right, s.t. '2' is above '3'. How many dots
    are placed depends on the SU($n$) under consideration}
 \label{fig:eytfig1}
\end{figure}

\emph{Rule, original version}--- %
  Let $T$ be a standard YT of size $N$. By sliding them to the right 
  where necessary, arrange all boxes of $T$ such that in each column 
  of the resulting \emph{extended} tableau the numbers in the boxes are 
  in sequence (\ie $i$ above $i$+1 above $i$+2 etc.). This will often require 
  leaving empty spaces between boxes (see Fig.~\ref{fig:eytfig1}). Mark each 
  by a dot. 
  To each dot $i$ we assign a number $a_i$ in such a way that the average
  of all $a_i$ within one column equals the average of the numbers in all boxes 
  in that column, where the $a_i$ have integer or half-integer values with a 
  spacing of $1$ between the numbers from one column.

\emph{Haldane--Shastry model.}---%
This version of the rule betrays the origin of the procedure: it comes from the physical 
problem of the Haldane-Shastry spin chain~\cite{haldane88,shastry88}, which consists 
of $N$ spins on a circle with a Heisenberg-type $J_{ij}\vco{S}_i\vco{S}_{j}$ interaction 
where the coupling $J_{ij}=|\eta_i-\eta_j|^{-2}$ decreases quadratically in the chord
distance.  
In the original model the $\vco{S}_i$ where $S=1/2$  SU(2) spins, but 
it has been generalised to fundamental irreps of SU($n$)~\cite{kawakami92prb1005,
kawakami92prb3191,ha-92prb9359,schuricht-06prb235105}.
This fully integrable model~\cite{talstra-95jpa2369} has a singlet ground state 
and the excitations are spinons, which can be thought of as delocalised domain walls 
('half a spin flip') in a background liquid with strong antiferromagnetic short range correlations.  
By interpreting each dot in the extended Young tableaux as a spinon, the eYT allow
obtaining the spinon content of the eigenstates (which also conserve total spin and total 
momentum) and moreover assign each spinon a momentum number $p_i$ connected to 
the $a_i$ from above via 
\begin{equation}
 p_i = 2\pi(a_i- 1/2)/N.
\end{equation}
The total momentum of a state is obtained by summing over all individual spinon momenta 
while the energy is essentially the sum of the squares~\cite{greiter-02jltp1029}. 
In both quantities we need to include the constant offset $p_0$ and $p_0^2$ respectively 
given by
\begin{equation}
  p_0 = \pi\frac{n-1}{n}N
  \label{eq:p0}
\end{equation}

The original rule is a succinct formulation of the basic idea, but it is not well suited to 
implementation on a computer, nor does it generalise directly to higher SU($n$) irreps.
If one wishes to use extended YT for computing eigenvalues of $C_N$ on tensor product
spaces of higher SU($n$) irreps, it is better to make use of the branching rule. 
Let us therefore reformulate the building rule.

\begin{figure}
  \centering
   \includegraphics[width=.528\textwidth]{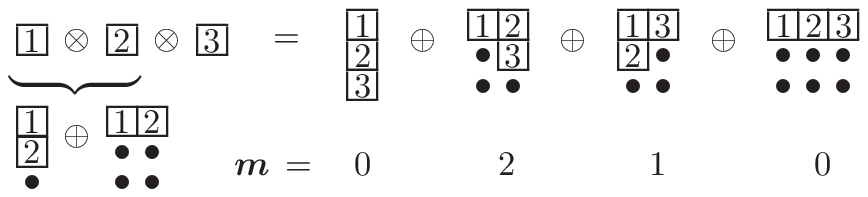}
  \caption{Building extended tableaux box-by-box.
           Since we consider SU(3), three rows are marked with dots.
           However, the momenta assigned do not depend on $n$}
 \label{fig:eytfig_branch}
\end{figure}

\emph{Rule, new version} ---%
  Given a standard YT $T$, we start with an incomplete extended tableaux 
  $E_1(T)$ containing only the single box labelled '1'. we then add a box labelled
   '2' to $E_1$ by looking whether in $T$, '2' appears in the first or second row. 
  In the former case we add '2' to the right of '1', while in the latter case
  we put it below '1'. This yields an, \ig still incomplete, 2-box extended 
  tableaux which we call $E_2$.
  We go on building the full extended tableau $E(T)$ step-by-step.
  In step $k$, having constructed the extended tableaux $E_k$, 
  we obtain the next one, $E_{k+1}$, in a similar way as $E_2$: 
  we look up in $T$ the row index $r_{k+1}$ of the box 'k+1' and compare 
  it to the one of 'k', which is $r_k$. If now $r_{k+1}\leq r_k$, we add a 
  new column at the right side of $E_k$ and put the box 'k+1' in its 
  $r_{k+1}$th row. Otherwise, \ie if $r_{k+1}>r_k$, we add 'k+1' into same 
  column as 'k', also in the $r_{k+1}$th row.
  The resulting extended tableaux we call $E_{k+1}(T)$. 
  After a total of $N-1$ additions we thus arrive at the final tableaux 
  $E(T)=E_N(T)$.

This procedure is now directly implementable and above all,  generalises  to 
products of higher (symmetric) representations. 
To compute the total momentum $p_T$ associated to the tableaux $T$,  we combine 
the spinon momentum numbers $a_i$ in each column $c$ of the extended tableau 
$E(T)$ into one \emph{column number} $b_c$
\begin{equation}
   b_c = \sum_{i\in c} \left(a_i -\frac{1}{2}\right)
\end{equation}
Written directly in terms of the average of the box-labels $\braket{i}_c$
and the number of boxes $k_c$ in the column, the $b_c$ are:
\begin{equation}
  b_c=\left(n-k_c\right)\left(\braket{i}_c-1/2\right) 
  \label{eq:bcnumbers}
\end{equation}
the momentum $p_T$ associated with $T$ is then simply 
\begin{equation}
 p_T = \frac{2\pi}{N} \frac{1}{n} \left(b_0+\sum_{c\in E(T)} b_c\right)
 \label{eq:momnumber}
\end{equation}
The momentum offset $b_0:=-(n-1)N^2/2$ is still necessary to 
ensure that the sum~(\ref{eq:momnumber}) is a multiple of $n$. 
Thus no matter which SU($n$) a tableau T pertains to (although clearly 
one must have$n>$\#rows in $T$), it is always assigned the same
momentum by our procedure.

\begin{figure}[t]
 \centering
  \includegraphics[width=.54\textwidth]{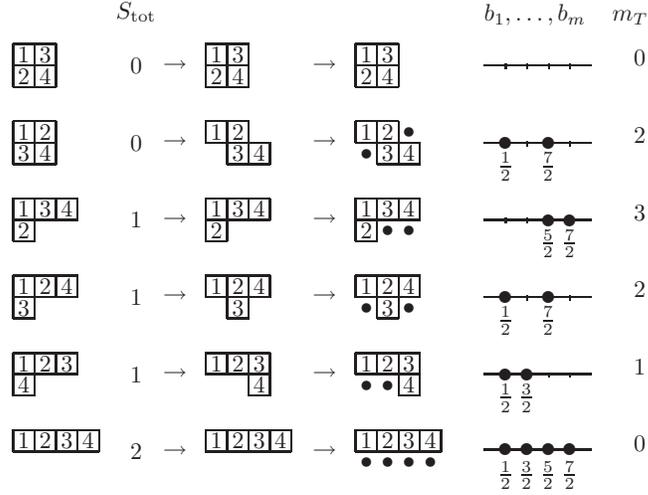}
 \caption{The complete list of extended Young tableaux for $N=4\times (S=1/2)$ 
          spins of SU($2$), $m_T$ is the integer momentum number $4p_T/2\pi$.
          }
 \label{fig:eytfig2}
\end{figure}

In Fig.\ref{fig:eytfig2} we show as an example the extension procedure for all total 
spin multiplets of the tensor product space $\left(\frac{1}{2}\right)^{\otimes 4}$ of four 
$S=1/2$. Since this is the tensor product of a fundamental representation, the shapes 
of the YT immediately tell us what irreps of $S_4$ the total spin multiplets belong
to. It is thus easily verified, that our procedure gives the correct values: The lone 
quintet $S^\tot=2$ must be fully symmetric and has therefore momentum $0$, the 
three triplets $S^\tot=1$ form the standard representation of $S_4$ (of dimension 3 
and associated with the partition $(3,1)$) while the two singlets belong to the self-
conjugate irrep $(2,2)$ (2 dimensional).

The Haldane-Shastry model is valid for $\vc{S}_i\in SU($n$)$ not only for $n=2$
and the mechanism of constructing excitations remains the same.  Therefore this 
connection of extended YT and HSM eigenstates exists not only for SU(2) but higher 
$n$ as well and since extended YT correctly describe the eigenstates of the Haldane-
Shastry model in all cases, it provides the strongest argument in favour of the 
correctness of our procedure.  
A rigorous mathematical proof would still be desirable however.

\begin{figure}[t]
  \centering
  \includegraphics[width=.54\textwidth]{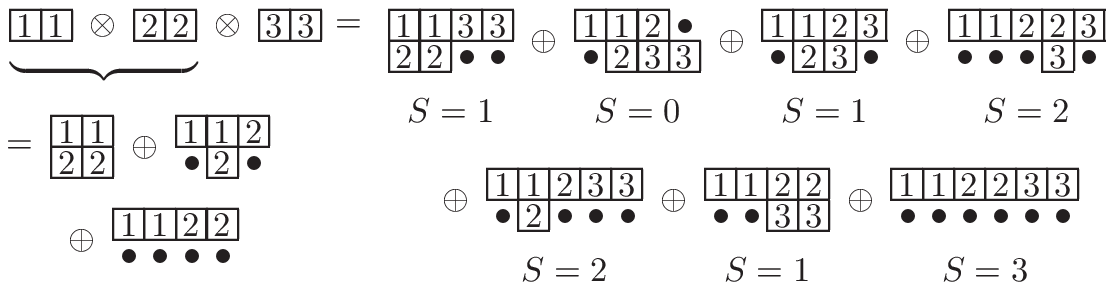}
  \caption{Higher representation extended YT can be built box-wise as well. 
          Ambiguity below which of two identical $k$-boxes to place a 
          $(k\!+\!1)$-box requires an additional rule}
 \label{fig:eytfig_hbranch}
\end{figure}

\section{Higher Representations \label{sec:high}}
While the spins in the Haldane-Shastry model can transform like any
fundamental SU($n$) representations, not just SU(2), a straightforward
generalisation to higher representations runs into difficulties. One
can nonetheless hope, that the \emph{mathematical} statement remains
valid for a suitable generalisation of the rule, which is what we
investigated. It turns out, that for products of symmetric SU($n$)
representations $V_\sigma$ (represented by single row YT with 
$|(\sigma)|=\sigma$ boxes) the box-by-box approach to building extended
tableaux generalises almost directly (see Fig.~\ref{fig:eytfig_hbranch}). 

We have to introduce only one additional condition coming from an ambiguity 
in where to put a box '$k\!+\! 1$' if there are several eligible open columns 
with boxes '$k$'. 
Given a tableau $T$ representing a multiplet in $V_\sigma^{\otimes N}$ 
(\ie each number appears $|\sigma|$ times) we require that the resulting 
extended tableau $E(T)$ is \emph{minimal}, \ie has as few dots/empty 
spaces as possible. 
The way to achieve this is to consider the boxes with number '$k\!+\! 1$' 
\emph{in increasing order of their row index in $T$}, \ie place higher boxes first. 

\begin{figure}[b]
  \centering
  \includegraphics[width=0.571\textwidth]{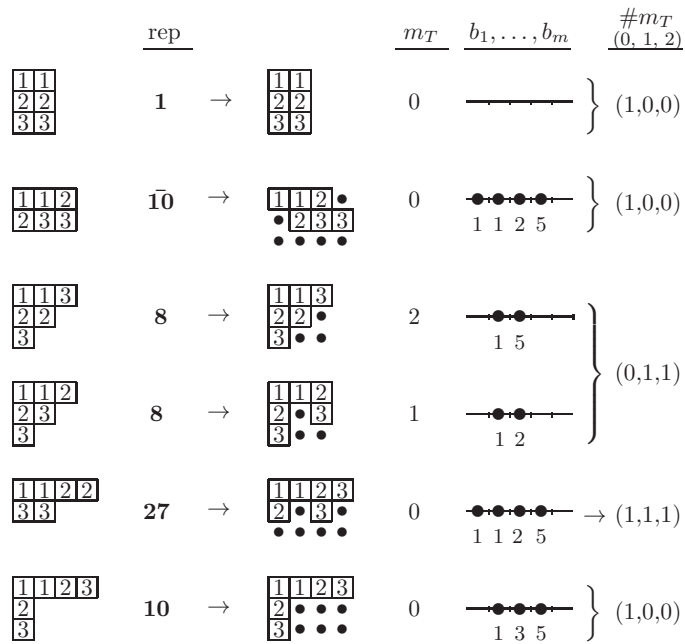}
  \caption{Here are some of the extended YT we find when combining
    $N=3$ rep $\bm{6}$ ($\hat{=}(2,0)$) of SU(3). 
    The last column gives the tally of momenta for the shape.  }
  \label{fig:eytfig_high}
\end{figure}

Consider the fourth YT from the top in Fig.~\ref{fig:eytfig_high}.
Assume we have already placed all '1's and '2's, giving us an (incomplete) 
extended tableaux with three columns, the first containing '1' above '2', in the 
second and third a '1' and a '2' respectively in the first row. If we now were to 
place the '3' from the lowest row first, then we have two eligible columns (the 
first and third). Depending on our decision, we would end up with two 
different extended tableaux, one with momentum $p=0$ the other with $p=1$. 
Placing the higher '2' (the one from the second row) first, there is no ambiguity, 
and we identify the extended YT with $p=1$ as the correct minimal one.

The only further change is a that the momentum offset $b_0$ 
acquires a factor $|\sigma|$:
\begin{equation}
  b_0(\sigma) = -\frac{n-1}{2} N|\sigma|
  \label{eq:offsethigh}
\end{equation}
The $|\sigma|$ dependence arises naturally as is explained in Appendix C.

\begin{figure}[t]
  \centering
  \includegraphics[width=0.58\textwidth]{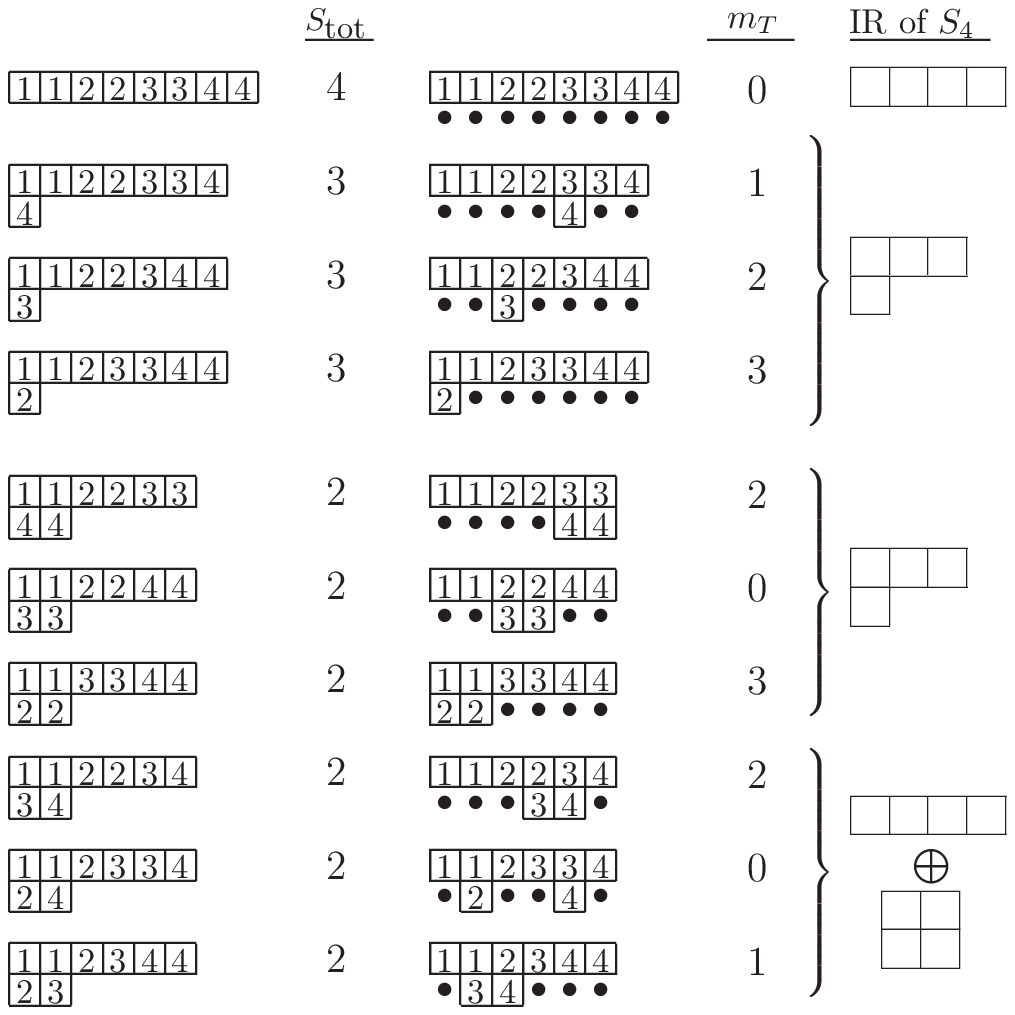}
  \caption{The complete decomposition of $V={S=1}^{\otimes 4}$ for $S_\tot\geq 2$ 
    and momenta assigned by our method. The right column shows
    the ad-hoc identification of YT with the irreps of $S_4$ as described in
    the text.}
  \label{fig:eytfig_N4S1}
\end{figure}

Figures~\ref{fig:eytfig_high} and~\ref{fig:eytfig_N4S1} show examples
of the extension in the case of $|\sigma|=2$.
In the absence of a rigorous mathematical understanding \emph{why} it works
and without the physical interpretation that backs up the extended YT procedure
in the case of fundamental SU($n$) representations, we checked the statement 
numerically for several SU($n$) up to $n=4$ and $N=16$ and find it does 
give the correct eigenvalue distributions (see Table~\ref{tab:numcheck}).

\begin{table}[b]
  \renewcommand\arraystretch{1.3}
  \setlength{\unitlength}{5pt}
  \centering
  {\small
     \begin{tabular}{lcccr}
       group& rep.  &   shape $\sigma\;$& $N_\mathrm{max}$&dim($V^{\otimes N}$)\\
       \hline
       SU($2$)&    $S\!=\! 1/2\;\hat{=}(1)$                 &{\tiny\yng(1)}   &16   &   65536 \\
                       &    $S\!=\! 1\;\hat{=}(2)$                     &{\tiny\yng(2)}  &14    & 4782969 \\
       SU($3$)& $\underline{3}\;\hat{=}(1,0)$         &{\tiny\yng(1)}  &12    &  531441 \\
                      &$\underline{6}\;\hat{=}(2,0)$          &{\tiny\yng(2)}   & 9    &10077696 \\
                      &$\underline{10}_3\;\hat{=}(3,0)$   &{\tiny\yng(3)}   & 7    &10000000 \\
       SU($4$)&$\underline{4}\;\hat{=}(1,0,0)$      &{\tiny\yng(1)}   &10   & 1048576 \\
                      &$\underline{10}_4\;\hat{=}(2,0,0)$&{\tiny\yng(2)}   & 7    &10000000 \\
    \end{tabular}
  }
  \caption{Unitary groups SU($n$) and maximal tensor powers $N$ for which we 
           verified the correctness of extended Young tableaux}
  \label{tab:numcheck}
\end{table}

Figure~\ref{fig:eytfig_N4S1} illustrates a limitation of our method.
As mentioned, multiplet subspaces $V_{\lambda}^{\oplus a_{\lambda}}$ in $V_\sigma^{\otimes N}$
comprise in general more than one irreducible representation of $S_N$.
Only for products of fundamental representations ($|\sigma|=1$) do
irreps of $S_N$ and SU($n$) coincide.
The $a_{S=2}=6$-dimensional subspace of all quintets in $(S=1)^{\otimes 4}$ 
for instance (\ie $V_{S=2}^{\oplus a_{S=2}}$) contains the three $S_4$ irreps
(3,1), (2,2) and (4). 
If we try an ad-hoc identification of these irreps with the YTs based on
the latter's structure (see Fig.~\ref{fig:eytfig_N4S1}), we see that the momenta
assigned by our procedure and the momenta one would expect from this 
identification do not match: the three tableaux we identified as belonging to 
irrep $(3,1)$ are assigned momenta 2,3 and 0, while one would expect 1,2 and 3.

Thus, while extended YT do produce the correct frequencies of momenta
for each subspace $V_{\bm{\lambda}}^{\oplus a_{\bm{\lambda}}}$
\emph{as a whole}, they give no help in identifying the $S_N$ irrep
content of multiplet subspaces (beyond what the momentum frequencies
themselves already reveal).

\section{Fast tableaux generation\label{sec:comp}}
In this section we want to elaborate how working directly with the Young tableaux, 
allows a useful speed-up of $C_N$ eigenvalue computations in the limit of large
$N$ and $n$.

In the introduction, we already mentioned briefly two traditional computational 
methods for obtaining the eigenvalues of the cyclic subgroup generator $C_N$. 
They are character theory and diagonalisation of the matrix of $C_N$ on total weight
representations. 
Both begin by writing down a product-state basis $\mathscr{B}_{\bm{w}}$ of a total 
weight subspace $\bm{w}_\tot^\z=\bm{w}$. In the case of $(S=1)^{\otimes 4}$ of 
SU(2) for instance this would be the 10 dimensional space spanned by
\begin{equation}
  C_4^i\ket{1,1,1,-1}, C_4^j\ket{1,1,0,0}, C_4^k\ket{1,0,1,0}
\end{equation}
where $\,i,j=0..3,\,k=0,1$ and the cyclic permutation $C_4$ is applied to a state in the 
natural way. 
Clearly, these states form the basis of a representation of $S_4$. 

\emph{Character theory -}There are in fact two methods based on group characters: 
one working with the characters of $S_N$ and another, simpler one, using the characters 
of $\mathscr{C}_N$. 

The former, mentioned here for the sake of completeness, obtains the $S_N$ character
$\bm{\chi}_{\bm{w}}$ of the total weight representation spanned by $\mathscr{B}_{\bm{w}}$ 
(referred to it simply as \emph{the} representation $\mathscr{B}_{\bm{w}}$ from now on), 
\ie we compute the trace of the representation matrix of one element from each conjugacy 
class in $S_N$ and then decompose this compound character using the formula
$a_\lambda = \frac{1}{N!} \sum_{[P]} |[P]|\chi_\lambda(P)\chi_{\bm{w}}(P) $,
where the sum runs over all conjugacy classes $[P]\subset S_N$ where $P$ is 
some representative of the class and $|[P]|$ is its cardinality.
The eigenvalues of $C_N$ follow directly, since each irrep $\lambda$ comes 
with a fixed set of eigenvalues.

One arrives at more efficient way of using characters by realising that 
$\mathscr{B}_{\bm{w}}$ is also a representation of $\mathscr{C}_N$, which means 
we can apply character decomposition directly to the representation matrices of
$C_N, C_N^2, \dots, C_N^{N}$. We can thus compute the multiplicity $f_m$ of a momentum 
number $m$ via the group characters of $\mathscr{C}_N$:
\begin{equation}
  f_m = \frac{1}{N}\sum_{k=1}^N \exp\left[\frac{2\pi\,\mathrm{i}}{N}\,m k\right]\; \mathrm{Tr}\,C_N^k
\end{equation}
where $m=0,\dots,N-1$ labels the irreducible representations 
and $k=1,\dots,N$ the classes in $\mathscr{C}_N$ and $\mathrm{Tr}\,C_N^k$ is
the trace of the $n_{\bm{w}}\times n_{\bm{w}}$ representation matrix
of the $k$th power of $C_N$.

This is both faster than the full character decomposition and does not
assume prior knowledge of all the irreducible $S_N$ characters (which would
in practice have to be computed too). We do however have to generate all 
the powers $C_N^k$ of $C_N$, which takes (at least) $O(N\,n_{\bm{w}})$ steps. 

We are not done yet however, for remember that the total weight representation $\mathscr{B}_{\bm{w}}$ 
contains not only the irrep $\bm{w}$ but also some with highest weight $\bm{w'}>\bm{w}$, 
which we need to sift out. 
The SU(2) case is straightforward: we simply run the procedure twice, once for $S_\tot^\z=S$ 
and once for $S_\tot^\z=S\!+\!1$ and then subtract the $S_N$-momentum tally of the latter 
from that of the former. 
The general case requires more work however: first, we need to know the positive integers 
$c_{\bm{w}\bm{w'}}$ recording how many states a representation $\bm{w}'$ contributes 
to $\bm{w}$.
The fact that only $\bm{w'}>\bm{w}$ contribute and $\bm{w}$ is contained exactly once 
means that viewed as a matrix, $\left( c \right)$ will be upper triangular with only 1s on 
the main diagonal.
To obtain the tally of all momenta for $\bm{w}^\tot$, we then need to take linear combinations 
of a certain number $h_{\bm{w}}$ of rows of this matrix, such that, all contributions of higher 
multiplets are cancelled.

Thus, total asymptotic complexity is
\begin{equation}
 \mathrm{C}_{\mathrm{char},C_N}= \myO{N\,h_{\bm{w}}\,n_{\bm{w}}}
 \label{eq:comp_char2}
\end{equation}
This is however still not as good as the conceptually simple diagonalisation we will turn 
to next.

\emph{Diagonalisation.}---%
Diagonalisation is straightforward: we write out the representation  matrix of $C_N$ 
in the basis $\mathscr{B}_{\bm{w}}$ and diagonalise it. 
In general, diagonalisation is of (time) complexity $\myO{m^3}$ for an $m\times m$ matirx, 
but since we are dealing with permutation matrices (in each row and column all entries 
are zero except for exactly one '1'), $\myO{m}$ steps suffice. Like with the previous 
character methods, we will also obtain $C_N$ eigenvalues belonging to SU($n$)
irreps of higher highest weight which can be got rid of in the same way, 
incurring the same $h_{\bm{w}}$ factor.

In the end therefore, diagonalisation is faster than the based character methods
and if we assume that the representation matrix of $C_N$ can be written down 
in $O(n_{\bm{w}})$ steps the final time and space requirements are
\begin{equation}
 \mathrm{C}_\mathrm{diag} = \myO{h_{\bm{w}}\,n_{\bm{w}}}.
 \label{eq:comp_diag}
\end{equation}

\emph{Extension.}---%
The extension procedure on the other hand works directly with the $N_\lambda$ 
Young tableaux on a shape $\lambda$, assigning each a momentum number 
$m(T)=0,\dots,N-1$. 
The key to making it superior to the other methods, is that it is possible 
to combine YT creation, extension and momentum computation efficiently into one procedure.
\begin{figure}
  \centering
    \includegraphics[width=.528\textwidth]{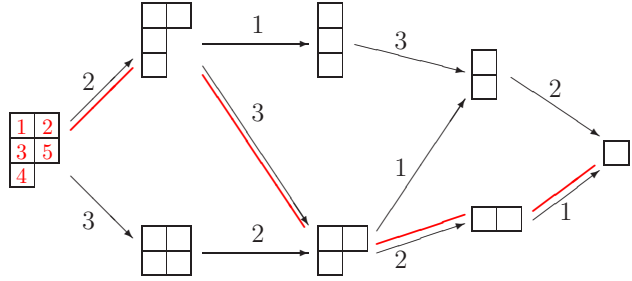}
  \caption{(Colour online) The standard YT branching graph of the shape $\lambda=(2,2,1)$. For 
                 all depth $j$ the arrow-labels denote into which line the index $N\!-\!j\!+\!1$ is to be put. 
                 Thus, the paths through the BG correspond 1-1 to all standard YT. An example of
                 such a path and the YT it corresponds to is shown in red.}
   \label{fig:bgraph}
\end{figure}

Let us first consider YT generation: one should exploit the branching property somehow, 
but a naive ansatz building up tableaux by adding box after box starting from scratch for 
each tableaux will require $\myO{N\,N_\lambda}$, which will only be marginally better 
than diagonalisation in the most interesting cases (the states with low total highest weight, 
\eg SU($n$) singlets) and due to the more intricate nature of the algorithms involved 
probably turn out to be somewhat slower in the less interesting ones (states with total 
highest weight close to the completely symmetric one).

We can achieve a complexity of $\myO{N_\lambda}$ however, if we store the branching 
information in a suitable way: the branching graph (BG). It encodes the relations between 
a shape $\lambda$ and all (valid) shapes $\mu \subset \lambda$ derivable from it by 
repeated regular removal of elementary shapes $\sigma$ in the form of a directed graph, 
where $\sigma$ was the shape associated with the SU($n$) irrep from which we build our 
product space.

The nodes of the graph are the shapes $\mu\subseteq\lambda$ (with $\lambda$ 
being the root) and a labelled edge $(\mu\rightarrow\nu;l)$ goes from shape $\mu$ to 
$\nu$ if and only if the latter can be obtained from  the former by a regular removal of 
one elementary shape $\sigma$. 
A regular removal is the inverse of a regular addition, which is defined as the addition 
of $|\sigma|$ boxes such that the resulting tableau is valid.
The label $l$ will be a list of length $|\sigma|$, recording into which row we put the first, 
second, third,\dots, $|\sigma|$th box. An example of a branching graph for the standard 
YT on shape $(2,2,1)$ is depicted in Fig.~\ref{fig:bgraph}. Since $|\sigma|=1$, the label 
consists of a single row-index only.

As long as $\sigma$ is a single-row tableau, as we always assume here, there will be at 
most one edge between nodes. However, the branching graph is also defined for tableaux 
built from multi-row-$\sigma$ (non-symmetric SU($n$) representations), but there it can 
happen that more than one edge leads from one node to the another (they will differ in 
their label however).  
Irrespective of the elementary tableau $\sigma$, each node in a BG can be assigned a 
depth, \ie a unique distance from the root, and it also holds that all branching graphs have
a unique lowest node (leaf) given by the elementary shape $\sigma$ itself.

Computing the branching graph of a compound shape $\lambda$ with $k$rows for some 
elementary shape $\sigma$ (where $|\lambda|=N|\sigma|$) requires
$\myO{\left(\begin{array}{c}k+|\sigma|\\
      k \end{array}\right)\,D_\lambda}$ steps, where
$D_\lambda^\sigma$ is the number of shapes $\mu$ obtainable from $\lambda$ 
by regular removal of $\sigma$. It can be estimated by (see Appendix B)
\begin{equation}
  D_\lambda^\sigma \leq \sum_{\lambda_1\geq j_1\geq\dots\geq j_k\geq 0}1=\left(\begin{array}{c} \lambda_1+k \\ \lambda_1 \end{array}\right)
 \label{eq:Dlambda_esti}
\end{equation}
The leading contribution is $D_\lambda^\sigma=\myO{N^k}$ (because
the first row $\lambda_1<|\sigma| N$) and thus we find that BG generation 
takes
\begin{equation}
  \mathrm{C}_\mathrm{BG}=\myO{k^{|\sigma|}\,N^k}
\end{equation}
time. We should point out that this is in general not polynomial, as it might appear at
first glance. Since $k$ is not independent of $N$, for \eg a square shape $N=k^2$ 
we indeed have $\mathrm{C}_\mathrm{BG}=\myO{\sqrt{N}^{|\sigma|}\,\exp[\sqrt{N}] }$. 
We will still profit from using the branching graph however, because even in these
cases $N_\lambda$ grows much faster still and thus dominates the total complexity
of computing the momenta (for more details see Appendix B).

One can now use the efficient graph iteration described in Appendix A to traverse all 
paths through the branching graph, simultaneously building up the extended YT as
we go. 
It is necessary to compute this both at once, because a modularized approach 
of extracting the paths first and then translating them one by one into extended YT 
incurs an additional $\myO{N}$ time factor coming from the fact that each path
is of length $N-1$.

The total asymptotic complexity (in both \emph{time} and \emph{memory})
achievable is therefore indeed determined purely by the number of YT
on $\lambda$
\begin{equation}
 \mathrm{C}_{\mathrm{extended YT}} = \myO{N_\lambda}
 \label{eq:comp_allYT}
\end{equation}

How much is this superior to diagonalisation? 
The biggest differences occur for low-weight SU($n$) representations (\eg singlets), 
and for these, the number of all multiplets $N_\lambda$ grows slower than $n_\lambda$, 
the size of the corresponding total weight space.

Take for instance $N$ spin $S=1/2$ ($N$ even): there are 
$\left(\begin{array}{c}N\\ N/2 \end{array}\right)$
$S_z^{\mathrm{tot}}=0$ states but only 
$$\left(\begin{array}{c}N\\ N/2 \end{array}\right)-\left(\begin{array}{c}N\\ N/2-1 \end{array}\right)= \frac{2}{N+2}\left(\begin{array}{c}N\\ N/2 \end{array}\right)$$ 
$S^{\mathrm{tot}}=0$ singlets.

In addition, the other methods incur the factor $h_{\bm{w}(\lambda)}$ because they 
need to repeated for higher weights, as described above. This factor, while trivial for 
SU(2), becomes increasingly important for larger $n$ .

\section{Conclusion \label{sec:conc}}
We have demonstrated how the extended Young tableaux method of
calculating the eigenvalues of $C_N$, the generator of the cyclic
subgroup $\mathscr{C}_N\subset S_N$, can be used not only for product
spaces of fundamental SU($n$) representations (associated with a
single-box Young diagram), but for those of higher ones as well, if
they are symmetric, \ie correspond to single-row diagrams.

Furthermore, since extended Young tableaux derive directly from the YT
on a shape $\lambda$, it is possible by exploiting the branching rule to 
speed up the computation of $C_N$ eigenvalues on 
$V_\lambda^{\otimes a_\lambda}$ to an asymptotic complexity $\myO{N_\lambda}$ 
in time and memory.

BS was supported by the Landesgraduiertenf\"orderung
Baden-W\"urttemberg.

\appendix

\section{Efficient graph iteration}
The branching graph (BG) is the key data-structure for implementing fast
Young tableaux generation, but the form presented above is not yet
sufficient to allow an efficient iteration over all paths through
it. We need to both add some additional information to the it and 
use suitable data-structures to guide the iteration itself.

To illustrate: Taking the BG as it is, we could for instance perform a
depth-first iteration: we use a size $N-1$ (= max. depth) array $c[.]$
to record to which child we descend to from the node $\mu[c](d)$ at
the depth $d$. In each step we then descend one level further down the
graph until we reach the leaf and there, having found a new path from root 
to leaf, we add it to our result and backtrack to the closest node 
where we can descend in a different direction.  
If we are only interested in the extended YT corresponding to the path
we can built it as we descend down to the leaf, and store it instead of 
the path. 

The problem is however, that in all this we descend and backtrack 
step by step through the graph, which will take on average $\myO{N}$ 
steps and this brings the total complexity up to $\myO{N\,N_\lambda}$.
\begin{figure}[b]
 \centering
    \includegraphics[width=.528\textwidth]{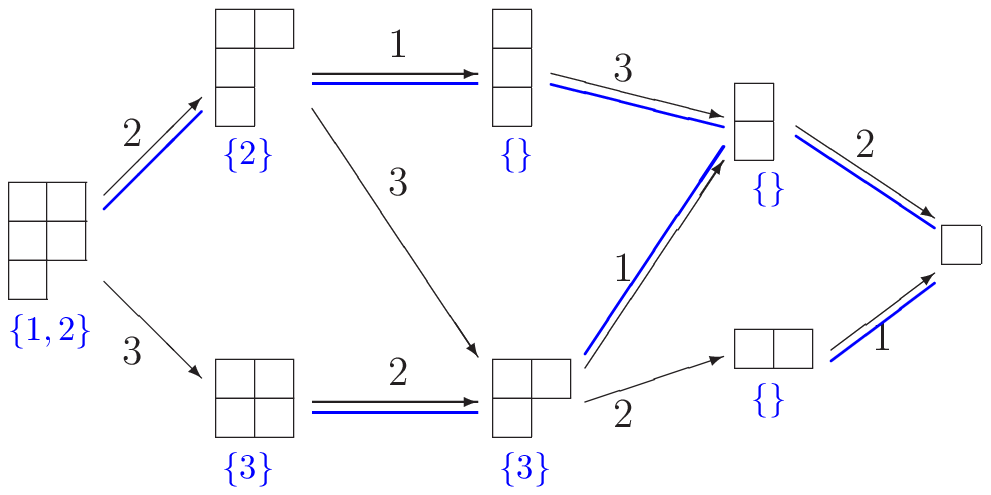}
 \caption{(Colour online) To enable efficient graph iteration, we need to augment
   the basic branching graph with additional information (shown in blue): to 
   each node we attach the potential backtracking-positions, i.e. all nodes with
   more than one child, lying on the \emph{leftmost path} descending from that 
   node to the leaf.}
 \label{fig:bgraph_enh}
\end{figure}

To do better, we perform a pre-computation before the iteration
itself adding the following information to each node $\mu$: 
  from $\mu$ we follow the \emph{leftmost path} (lmp) to the leaf 
  and, while we descend, push on a \emph{stack} $K_\mu$ all the 
  nodes with more than one child, because only these will be potential
  candidates to backtrack to (see Fig.~\ref{fig:bgraph_enh}).  We also
  save $T_\mu$ the 'incomplete' extended YT containing all numbers 
  $N\!-\!d(\mu), N\!-\!d(\mu)\!-\!1,\dots,1$ where $d(\mu)$ is the depth of 
  node $\mu$.

Precomputation of the triple $K_\mu,l_\mu,T_\mu$ for all nodes of a BG
for shape $\lambda$ built from elementary shapes $\sigma$ takes
\begin{equation}
 \mathrm{C}_{\mathrm{pre}} = \myO{N}\myO{D_\lambda^\sigma} = \myO{N^{k+1}}
 \label{eq:precomp}
\end{equation}
where we used that the function $D_\lambda^\sigma$ counting the 
number of shapes $\mu\subset\lambda$ obtainable from $\lambda$ by 
regular removal of elementary tableaux $\sigma$ is bounded from 
above by $N^k$ (see Appendix B).
We see it is only linearly more demanding (in $N$) than computing the
basic branching graph .

Let us now sketch an iteration process which uses this precomputed 
information.
The ingredients are first the array $c[.]$, already known from the
naive iteration above and still needed to keep track of where we have
descended to from the node $\mu[c](d)$ lying at a depth $d$. Furthermore
we introduce a stack $K$ (the backtrack-stack) which will at all times
contain the depths of those nodes on our current path, where we could 
descend in a different direction, \ie those which have more than one 
child \emph{and} have not yet been exhausted ($c[d]<$~\#children of 
node $\mu[c](d)$).  
Flow control requires only a single 'while'-loop which is repeated 
as long as $K$ is nonempty.

The loop performs the following steps
\begin{itemize}
 \item[0.] assume we enter the loop with $c[.]$ initialised and a
complete extended YT $E$ (from initialisation or the previous pass)
 \item[1.] first, retrieve the uppermost element (depth) from $K$ (removing it
in the process)
 \item[2.] if that element is, say, $j$, increment $c[j]$ by one, reset
$c[i]=1$ for all $i>j$ and descend to $\nu:=\mu[c](j+1))$ (the next, still
unvisited child of node $\mu[c](j)$)
 \item[3.] update stack $K$: if $c[j]<$~\#children of $\mu[c](j)$
$\rightarrow$ push $j$ back onto $K$
 \item[4.] in any case: push all nodes (depths) indicated in the backtrack-list 
 of node $\nu$ onto $K$ (the blue lists in Fig.~\ref{fig:bgraph_enh})
 \item[5.] obtain next extended YT: drop the indices $1,\dots, j$ from $E$
          and join the remainder with the (incomplete) YT $T_{\nu}$ 
          which was added to node $\nu$ during the pre-computation
\end{itemize}

For standard YT, where the number of boxes $|\sigma|$ in the elementary
tableaux is equal to one, joining two parts of an extended YT can be done
in a single step:
we need only check wether $\mathrm{row}(j)<\mathrm{row}(j\!+\!1)$. 
If so, we merge the leftmost column of the remainder with the rightmost column 
of $T_{\nu}$. If not, we simply concatenate. 
If we are considering tableaux with $|\sigma|>1$, joining requires more
steps, but can always be achieved in $\myO{|\sigma|}$ time.

In all, the algorithm sketched above needs $\myO{1}$ (or
$\myO{|\sigma|}$ i.g.) steps to generate one path/extended Young
tableaux, and thus the full iteration requires $\myO{|\sigma|\,N_\lambda}$ 
time and $\myO{N\,N_\lambda\,|\sigma|}$ memory if we store the complete
list of extended tableaux. If we only keep the momenta, $\myO{N_\lambda}$
memory will suffice.

\section{Branching graph size}
The time required to generate the basic branching graph for a shape
$\lambda$ built up from $N$ elementary tableaux $\sigma$ as well as
augmenting it in preparation for efficient iteration is determined
mostly by its size, i.e. the number of its nodes. This in turn is just
$D_\lambda^\sigma$, defined as
\begin{eqnarray}
  D_\lambda^\sigma & = &\#\textrm{diagrams }\mu\textrm{ with }\mu\subset\lambda\textrm{ and }\mu\textrm{ obtained from } \nonumber\\
                   &  &\lambda\textrm{ by regular removal of one or more shapes }\sigma
 \label{def:Dlambda}
\end{eqnarray}
Our goal is now to find a good estimate for $D_\lambda^\sigma$.\\
Defining $D_\lambda:=D_\lambda^{\sigma=(1)}$, we can use it as an upper bound
on $D_\lambda^\sigma$, as the additional $\sigma$-dependent
constraints in (\ref{def:Dlambda}) serve only to \emph{decrease}
the number $\mu$ that are compatible.

But $D_\lambda$ is easily expressed as the multiple sum
\begin{equation}
  D_\lambda=\sum_{j_1\leq\lambda_1}\sum_{j_2\leq\mathrm{Min}(\lambda_2,j_1)}\dots \sum_{j_k\leq\mathrm{Min}(\lambda_k,j_{k-1})}\!\!\!\! 1
  \label{eq:Dlambda_bound}
\end{equation}
where $k$ is the number of rows in $\lambda$. 
This can be estimated from above by forgetting about Min($\lambda_i$,$j_{i-1}$) and 
bounding $j_i$ just by $j_{i-1}$ instead
\begin{equation}
 D_\lambda\leq \sum_{j_k\leq j_{k-1}\leq\dots\leq j_1\leq\lambda_1} \!\!\!\!\!\!\!1 \;= \left(\begin{array}{c} \lambda_1+k \\ \lambda_1 \end{array}\right)
  \label{eq:Dlambda_bound2}
\end{equation}
Two instances are of particular interest (set $\sigma=1$):
\renewcommand{\theenumi}{\alph{enumi}}
\renewcommand{\labelenumi}{(\theenumi)}
\begin{itemize}
\item $\lambda_1=N, k=1$ in which case the estimate gives almost the
      exact result ($D_\lambda = N$ compared to $\left(\begin{array}{c}
      N+1 \\ N \end{array}\right)=N+1$)
\item $\lambda_1= N/k=:m, k>1, k|N$(rectangle shape) where the above 
  	estimate gives the exact result, as we will show in the following.
\end{itemize}
Building the branching graph for a rectangular shape removing in turn 
$1,2,\dots,s,\dots, N\!-\! 1$ boxes is equivalent to building (box by
box) shapes with no more than $k$ rows and $m$ columns. Without
restrictions, the number of shapes with $N$ boxes is simply $p(N)$, the
number of integer partitions of N. With the restrictions, we must instead
use $p_{<k;m}(n)$, the number of integer partitions using at most $m$
summands of size $\leq k$. Thus we obtain the exact branching graph
size for a rectangular shape $\lambda = (m,\dots,m)$, if we sum this 
over all steps $s=1,\dots, N$:
\begin{equation}
 D_{(m,\dots,m)} = \sum_{s=1}^N p_{\leq k;m}(s)\; .
 \label{eq:Noverk}
\end{equation}
However, a little thought reveals that this and the sum~(\ref{eq:Dlambda_bound2})
are, in fact, the same, proving that in this case the bound~(\ref{eq:Dlambda_bound})
is tight and the number of nodes is exactly given by $(m+k)!/k!m!$.

What is the asymptotic complexity in terms of $N$? If $k\ll \lambda_1\approx N$
(or vice versa), clearly $(\lambda_1+k)!/k!\lambda_1!\leq (N+k)!/k!N!=\mathrm{O}(N^k)$
and therefore polynomial in $N$. However, if $k\approx \sqrt{N}$ (\eg shapes of square 
or triangular form like $(k,k\!-\! 1,\dots,1)$), then our upper bound is
$$D_\lambda = \myO{\frac{\Gamma(2\sqrt{N})}{\Gamma^2(\sqrt(N)) } } = \myO{\exp[\sqrt{N}] } $$
and since we have shown that it is tight in the case $\lambda_1=N/k$, we see
that there are indeed shapes for which computing the branching graph is of
nearly exponential complexity. But exactly these shapes also have the highest 
number of YT, growing like $\myO{\exp[N]}$, \ie fully exponential. 
Therefore computing the branching graph is always worthwhile, as it is in 
all cases much less costly than generating the YT from it.

For many other combinations of $N$ and $k$~(\ref{eq:Dlambda_bound})
overestimates the size of the the branching graph considerably. Take
$\lambda=(N-k+1,1,\dots,1)$. Assuming $N\!-\!k\!+\!1\!>\!k$ the true
value is $D_\lambda =
(\lambda_1-1)^2+(\lambda_1-1)(\lambda_1-k+1)\approx
2\lambda_1^2-k\lambda_1=\myO{\lambda_1^2}$ independent of k (as long
as it remains smaller than $\lambda_1$) while~(\ref{eq:Dlambda_bound})
yields $\myO{\lambda_1^k}$.

\section{Momentum offset $b_0$}
Given a tableaux $T$ built from $N$ elementary tableaux $\sigma$ we may 
interpret it as pertaining to the product space $V_\sigma^{\otimes N}$
of any SU($n$) where $n$ is at least as large as the number of rows in
$T$.
We want to show here, that the momentum assigned to $T$ via the sum 
~(\ref{eq:momnumber}) is independent of this interpretation, \ie independent
of $n$.
\begin{equation}
  \sum_{c\in E(T)}\! b_c = \sum_c (n-k_c)(\braket{i}_c-\frac{1}{2}) = n \left(\sum_c \braket{i}_c - \frac{1}{2}c_T\right) - \frac{1}{2} |\sigma| N^2
\end{equation}
where we defined $c_T$ as the number of columns of $E(T)$ and used the relations 
$c_T=\sum_c 1$, $\sum_c k_c = |\sigma|N$ and $\sum_c k_c\braket{i}_c = |\sigma|N(N+1)/2$.
As a reminder, $k_c$ is the number of boxes in column $c$ of the extended
tableaux $E(T)$ and $|\sigma|N$ is just the total number of boxes in $E(T)$
(and therefore also in $T$).

\begin{table}[t]
  \renewcommand\arraystretch{1.3}
  \centering
  {\small
 \begin{tabular}{cc@{$\,\Rightarrow\,$}c|c}
   $\quad\sum_c \braket{i}_c\quad$ &  $\quad N\quad$  &  $\quad c_T\quad$ & total \\
   \hline
     half-integer    &  even &  odd   & integer \\
       "                     &  odd   & even  &   "   \\
     integer           &  even & even   &   "   \\
       "                     &  odd   & odd    &   "   \\
 \end{tabular}
 }
  \caption{The four possible parity combinations of $\sum_c \braket{i}_c$ and $N$. 
           All lead to an integer value for the total momentum~(\ref{eq:b0derivation}).}
  \label{tab:pTsumcombos}
\end{table}

We see, that if we add the offset momentum number $b_0=-(n-1) N^2 /2$ we arrive at
\begin{equation}
  b_0 +\! \sum_{c\in E(T)}\! b_c = n\left(\sum_c \braket{i}_c - \frac{1}{2}(c_T + |\sigma| N^2) \right)
  \label{eq:b0derivation}
\end{equation}
and thus $n$ cancels when computing the momentum $p_T=2\pi/Nn (\sum_c b_c + b_0)$.

What still needs to be checked is whether the quantity in parenthesis in~(\ref{eq:b0derivation})
is always an integer. To see that this is indeed the case, we need to analyse 
the relationship between $N$, $c_T$ and $\sum_c \braket{i}_c$. Since in all columns of $E(T)$, 
the boxes are in sequence, $\sum_c \braket{i}_c$ is always either integer or half-integer. 
In fact we can express each summand as $\braket{i}_c=j_c+(k_c-1)/2$, where $j_c$ is number in the 
uppermost box. Therefore, $\braket{i}_c$ is half-integer, if
and only if there is an even number of boxes in column $c$. 
Now assume $\sum_c \braket{i}_c$ is half-integer. This means, we must have 
an \emph{odd} number of columns with an even number of boxes. If now $N$ is even, there is an 
even number of boxes left to be distributed over rows with an odd number of boxes in them. 
This means that this number of (odd-box-number) columns must be even. Thus in this 
case $c_T/2$ is half-integer while $N^2/2$ is integer and in total sum~(\ref{eq:b0derivation})
is of the form $n\times$integer.
It is not hard to see that in the other three cases this holds as well (see Tab.~\ref{tab:pTsumcombos}).\\

\bibliographystyle{unsrt}

\end{document}